\documentclass[aps,prb,twocolumn,showpacs,amsmath,amssymb]{revtex4}

\usepackage{graphicx}
\usepackage{epsfig}
\usepackage{subfigure}

\begin{document}

\title{Weak localization correction to the density of transmission eigenvalues
   in the presence of magnetic field and spin-orbit coupling for a chaotic
   quantum dot}

\author{B. B{\'e}ri}
\author{J. Cserti}
\affiliation{Department of Physics of Complex Systems,
E{\"o}tv{\"o}s University
\\ H-1117 Budapest, P\'azm\'any P{\'e}ter s{\'e}t\'any 1/A, Hungary}

\begin{abstract}

We calculated the weak localization correction to the density of 
the transmission eigenvalues in the case of chaotic quantum dots 
in the framework of Random Matrix Theory 
including the parametric dependence on
the magnetic field and spin-orbit coupling. 
The result is interpreted in terms of
spin singlet and triplet Cooperon modes of conventional diagrammatic
perturbation theory. 
As simple applications, we obtained the weak localization correction 
to the conductance, the shot noise power and the third cumulant of the 
distribution of the transmitted charge.

\end{abstract}


\maketitle

\section{Introduction}
\label{sec:intro}
Transport in a two dimensional electron gas is affected by the spin-orbit
coupling. The most common signature is the weak (anti)localization. It is a small
correction  to the conductance due to the interference of time reversed
trajectories. The sign of the correction depends on
the presence or absence of the spin-orbit scattering. The correction is suppressed if a
time reversal symmetry breaking magnetic field is present\cite{mesoreview1,SimonsAltshuler,Hikami,Bergmann}. The spin-orbit term
in the Hamiltonian has a form of a non-abelian vector potential\cite{MathurStone}. In the case
of quantum dots, if the spin-orbit coupling strength is position independent, a
gauge transformation can be done, which results in an effective Hamiltonian
with reduced spin-orbit coupling, and a rich variety of symmetry classes \cite{AF}.  If the spin-orbit coupling depends
on the position the transformation can not be done any more\cite{BCH,CBF}. As
a consequence, the accessible symmetry classes are the three standard classes of
Dyson, classifying the systems according to the presence or absence of time
reversal and spin rotation symmetry.   

For quantum dots with chaotic dynamics random matrix theory gives a convenient way to
describe the transport properties, provided that 
the electron transit time $\tau_{\rm erg}$ is much shorter than the other time scales of
the problem (mean dwell time $\tau_{\rm dw}$, spin-orbit time $\tau_{\rm so}$,
magnetic time $\tau_{\rm B}$, inverse level spacing)\cite{RMTQTR}. Constructing the appropriate RMT
models for the crossover between the symmetry classes, the magnetic field and
spin-orbit coupling dependence 
of the average conductance was calculated in
Refs. \onlinecite{AF,BCH,CBF}. The theoretical results are confirmed by
numerical simulations \cite{Jens} and they are in good agreement
with the experiments \cite{Zum02,Zum05}. 

If one would like to calculate the averages of other
transport properties, such as the shot noise power, higher order cumulants
of the distribution of the
transmitted charge, or any other linear statistics, the density of
transmission eigenvalues is needed \cite{RMTQTR}. 
Jalabert et al.\cite{JPB} gave the weak localization correction to the
transmission eigenvalue density for chaotic quantum dots belonging to Dyson's three
symmetry classes. Our work extends this result to the crossover regime between
these classes. We present a calculation of the
dependence of the weak localization correction to the transmission eigenvalue
density on spin-orbit coupling and perpendicular magnetic field. For the sake of simplicity we restrict our attention to
symmetric dots, i.e. we assume that the two leads attached to the cavity
support the same number of channels $N$, and for technical reasons we consider the
case of  $N \gg 1$. 

In some other sense
we complement the work done by Nazarov \cite{Nazarov95}, who calculated the crossover
behaviour of the weak localization correction to the transmission eigenvalue density for disordered
samples in dimensions $d=1,2,3$. Here we give the results for $d=0$
corresponding to a quantum dot.   

The study of the transmission eigenvalue density
is interesting not only because of the practical implications related to the
linear statistics, but it is instructive by itself as well, since it gives a
deeper insight to the weak (anti)localization phenomenon. In the cases of
 Dyson's symmetry classes the weak localization
correction to the transmission eigenvalue density is of the form of Dirac
delta peaks at the endpoints of the spectrum\cite{JPB}. Our analytical, closed-form
result shows  that these peaks broaden in the crossover regime, but the
correction still remains singular. Furthermore, similarly to higher
dimensional cases of Ref.~\onlinecite{Nazarov95}, it is possible to identify 
the peaks as originating from the singlet and triplet sectors of the Cooperon modes
of conventional diagrammatic perturbation theory. Our result also enables us
to study the transition from weak localization to weak antilocalization on the level
of transmission eigenvalues.

As applications, we calculate the weak localization correction
to the conductance, the shot noise power and the third cumulant of the
distribution of the transmitted charge. In the case of the conductance we
recover the result of Ref.~\onlinecite{BCH}, giving a verification of our
calculations. For the shot noise we find that for the symmetric cavities studied in this paper
the weak localization correction is absent in the full
crossover regime.  The third cumulant of the transmitted charge behaves the
opposite way. It is ``crossover induced'' in the sense that the classical
contribution vanishes \cite{blanter01,nagaev02-2} and  the weak localization term is nonzero only in the crossover regime.  

The paper is organized as follows. In the next section we specify the systems
under consideration and the model applied for the RMT description. We briefly
summarize the formal definition and the practical importance of the density of
transmission eigenvalues. In Sec. \ref{sec:result} we present our main result, the weak localization
correction to the transmission eigenvalue density, and analyze its behaviour
as a function of the degree of time reversal and spin rotation symmetry breaking. In Sec. \ref{sec:applications} we
apply  our result to the transport properties above.  Finally we conclude in Sec.~\ref{sec:concl}.

\section{Description of the systems and the RMT model}
\label{sec:sysmod} 

Let us consider a chaotic quantum dot with two leads attached to it. We
assume, that the number of propagating modes is the same for both leads. We choose the
spin-orbit coupling to depend on the position to avoid the reduction of the
coupling strength. The weak magnetic field is
perpendicular to the plane of the dot. The assumptions for the spin-orbit
coupling and the magnetic field ensure that  our system exhibits
a crossover between Dyson's standard symmetry classes.

The transmission eigenvalues are the eigenvalues of the matrix product
$t't'^{\dagger}$. Denoting the number of modes in a lead with $N$, the
transmission matrix $t'$, describing the transmission from lead 2 to lead 1, is an $N \times N$ matrix with quaternion
elements. It is submatrix of $S$,  the $2N \times 2N$ scattering matrix of the system:
\begin{equation}t'=W_1 S W_2,\label{eq:tprime}\end{equation}
where $W_1$ is an $N\times 2N$ matrix defined by $(W_1)_{ij} =1$ if $i=j$  and $0$ otherwise, $W_2$ is a $2N\times N$ matrix with $(W_2)_{ij} =1$ if $i=j+N$  and $0$ otherwise. The product $t't'^{\dagger}$ has $2N$
eigenvalues, where the factor two comes from the quaternion structure of the
matrix elements. If the system is time reversal invariant, there are $N$
twofold degenerate levels.

We assume that the system can be described with random matrix theory,
i.e. $\tau_{\rm erg} \ll$ $\tau_{\rm dw}$, $\tau_{\rm B}$, $\tau_{\rm so}$,
where the magnetic time is related to the flux $\Phi$ through the system as \cite{AF,FrahmPichard}
\[\frac{1}{\tau_{\rm B}}= \frac{\kappa}{\tau_{\rm
    erg}}\left(\frac{\Phi}{\Phi_0}\right)^2, \]
where $\Phi_0$ is the flux quantum and $\kappa$ is a numerical factor of order
unity.

To give a statistical description of  the crossover behaviour of
the transmission eigenvalues we need an RMT model for the scattering matrix in
the crossover regime. This is provided by the ``stub model'' 
\cite{WavesRM} which was adapted for the system under consideration in
Ref.~\onlinecite{BCH}. In this approach the S matrix is represented as 
\begin{equation}
S=PU(1-RU)^{-1}P^{\dagger },  \label{eq:SU}
\end{equation}
with
\[R=Q^{\dagger }rQ.\] 
In the above expression $U$ is an $M \times M$ random
unitary symmetric matrix taken from Dyson's
circular orthogonal ensemble\cite{RMTQTR} (COE) and
$r$ is a unitary matrix of size $M-2N$. The $2N \times M$ matrix $P$
and the $(M-2N) \times M$ matrix $Q$ are projection matrices with $P_{ij}
= \delta_{i,j}$ and $Q_{ij} = \delta_{i+2N,j}$. The quaternion elements
of the matrices $U$, $P$, and $Q$ are all proportional to the 
$2 \times 2$ unit matrix $\openone_2$. The matrix $r$
is given by
\begin{equation}
  r ={\rm e}^{-{i \over M} H'},
\end{equation}
where $\Delta$ is the mean level spacing of the dot.  
$H'$ is an $(M-2N)$ dimensional quaternion matrix generating the
perturbations to the dot Hamiltonian, 
\begin{eqnarray}
  H' &=& i x X \openone_2   + i a_{\rm so} (A_1 \sigma_x +
  A_2 \sigma_y). \nonumber \\
  \label{eq:dH}
\end{eqnarray}
Here $A_j$ ($j=1,2$) and $X$ are real antisymmetric matrices of 
dimension $M-2N$, with $\mbox{Tr}\, \langle A_i A_j^{\rm T}\rangle = M^2 \delta_{ij}$
and $\mbox{Tr}\,\langle X X^{\rm T}\rangle = M^2$ and $\sigma_i$ are the
 Pauli matrices. The first term in
\eqref{eq:dH} describes the time reversal symmetry breaking through the magnetic
field. The second term, having a symplectic symmetry,
corresponds to the Rashba and/or Dresselhaus terms in the case of position
dependent spin-orbit coupling\cite{tauso}. The dimensionless parameters $x$ and $a_{\rm
  so}$ are related to the corresponding time scales as 
\[ x^2 = \frac{2  \pi \hbar}{\tau_{\rm B} \Delta},\quad a_{\rm so}^2 =\frac{ 2  \pi \hbar}{\tau_{\rm so} \Delta}.  \]
At the end of the calculation the limit $M \rightarrow \infty$ should be taken.

The density of transmission eigenvalues is defined as
\begin{equation}\begin{array}{ll}
\rho(T)&=\left\langle\sum_i  \delta(T-T_i)\right\rangle\\
&=-\frac{1}{\pi}\lim_{\epsilon\rightarrow 0^+}\
  \rm{Im}\left\langle\rm{Tr}\frac{1}{T-t't'^\dagger+i\epsilon}\right\rangle\end{array}\label{eq:densgen},\end{equation}
where the trace is taken over channel and spin indices. Having $\rho(T)$ at
hand we can calculate the ensemble  average of any linear statistics \cite{RMTQTR}
\[A=\sum_{i=1}^{2N} a(T_i), \]
as
\[\langle A \rangle =\int \rho(T)a(T)dT.\]

Prominent examples for linear statistics are the conductance, the shot noise
power, or the cumulants of the distribution of transmitted charge \cite{RMTQTR,BB,LLY95}. The
weak localization correction for the linear statistics can be obtained from
the weak localization correction to the transmission eigenvalue
density.

\section{Result and discussion}
\label{sec:result}
To find the density of transmission eigenvalues, one has to substitute the scattering matrix
\eqref{eq:SU} into the definition \eqref{eq:densgen} 
using \eqref{eq:tprime}, expand the inverses and calculate the average
with the help of the diagrammatic technique of Ref.~\onlinecite{diagrams} up to subleading
order in the small parameter $1/N$. The details of the calculation can be
found in Appendix~\ref{app:details}. The result is
\[\rho(T)=\rho_0(T)+\delta \rho (T). \]

With the $O(N)$ contribution  
\begin{equation} \rho_0(T)=\left\{ \begin{array}{ll}
\frac{2N}{\pi \sqrt{T(1-T)}} & \textrm{if $0< T< 1$} \\
0 & \textrm{otherwise} \\
\end{array} \right., \label{rho0}\end{equation}
we recover the known result of
Refs.~\onlinecite{JPB,BarangerMello,NazarovR,RMTQTR}.
The factor of two accounts  for  the spin, as a consequence of the known fact, that the
$O(N)$ term is insensitive to the spin-orbit interaction and the
magnetic field.

The main result of our paper is the closed-form, analytical expression for the weak
localization correction to the density of transmission 
eigenvalues,
\begin{equation}\begin{array}{c}\delta \rho(T) =\\
\quad\\
 \frac{1}{\pi \sqrt{T(1-T)}}\sum\limits_{n=0,1}(-1)^n
  \left(\frac{\Gamma_1(y_n)}{2}+\frac{\Gamma_2(y_n)-\Gamma_3(y_n)}{4}
  \right)\end{array}\label{deltarho} \end{equation}
 for $0<T<1$ and $0$ otherwise. The variables $y_n$ are $y_0=T$ and $y_1=1-T$. 
The expression for $\Gamma_m$ is
\[ \Gamma_m(y)=\frac{\gamma_m(\gamma_m+2N)}{\gamma_m^2+4N(\gamma_m+N)y},\]
with 
\[ \gamma_1=a_{\rm so}^2+x^2,\quad \gamma_2=2a_{\rm so}^2+x^2,\quad
\gamma_3=x^2.\]
Note that $\delta \rho (T)$ is antisymmetric with respect to the point
$T=1/2$.

 As a verification of \eqref{deltarho}, we consider the limits of Dyson's symmetry
 classes. Labeling them in the usual way with the index $\beta$, they correspond to
  $x^2,a_{\rm so}^2 \ll N$ for $\beta=1$ i.e. systems with time reversal
 and spin rotation symmetry, $x^2 \gg N$ for $\beta=2$, that is, systems where the
 time reversal symmetry is broken by the magnetic field and  $ x^2 \ll N \ll a_{\rm so}^2$ for
 $\beta=4$, for time reversal invariant systems without spin rotation
 symmetry. In these limits expression \eqref{deltarho} recovers the known result 
\begin{equation} \label{eq:drhopure}
  \delta \rho(T) = 2{2 - \beta \over 4\beta} \left[ \delta(T-0^+) - \delta(T - 1+0^+) \right]
\end{equation}
of Refs. \onlinecite{RMTQTR,JPB}. The factor of two in
\eqref{eq:drhopure} is due to the twofold degeneracy of transmission
eigenvalues for $\beta=1,4$,  mentioned in Sec \ref{sec:sysmod}.

To get some insight how the weak (anti)localization peaks in
\eqref{eq:drhopure} emerge, let us have a closer look at $\delta \rho(T)$ as
it approaches the different limits. 
For $\gamma_m \ll N$ the expression for
$\Gamma_m(y)$ is well approximated by a Lorentzian in the variable $\sqrt{y}$,
\[\Gamma_m(y) \approx \frac{\gamma_m/2N}{(\gamma_m/2N)^2+(\sqrt{y})^2}.\]
In the opposite limit,  $\gamma_m \gg N$, the functions $\Gamma_m$ become
independent of $y$. Thus going to $\beta=1$, $\Gamma_2$ and $\Gamma_3$ cancel, and
$\Gamma_1$, together with the inverse square root prefactor evolves to the
peaks in \eqref{eq:drhopure} at the edge of the spectrum. Close to
$\beta=2$  the correction vanishes. Notice, that $\Gamma_3$ is independent of
the spin-orbit coupling parameters. Specially in zero magnetic field it always
gives Dirac delta contributions at $T=0$ and at $T=1$. Approaching $\beta=4$,  
the contributions from $\Gamma_1$, $\Gamma_2$ disappear, thus the zero magnetic field peaks
associated with $\Gamma_3$ show up as the weak antilocalization
correction.

\begin{figure}
\epsfig{file=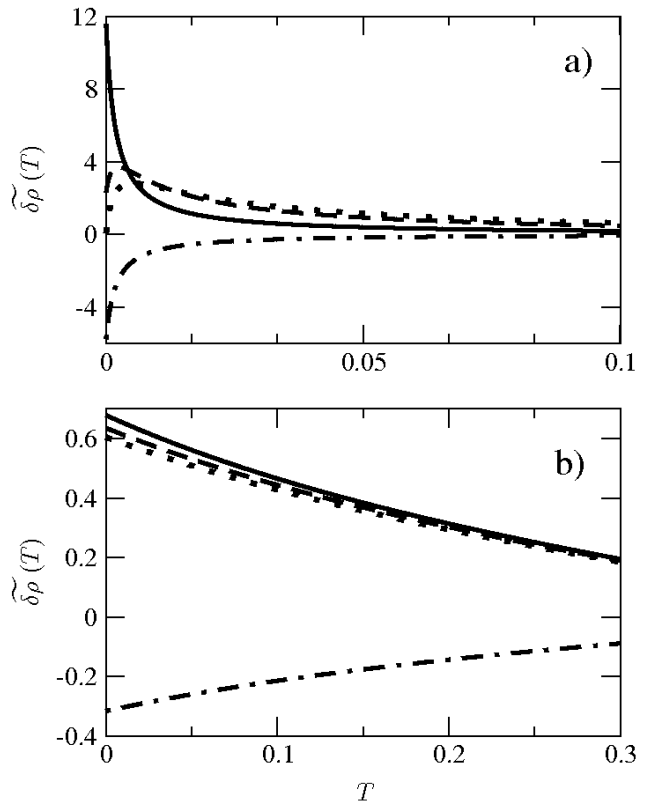,height=11cm} \\
\caption{The weak localization weak antilocalization transition for two values
  of the magnetic field, $x/\sqrt{N}=0.3$ (a) and $x/\sqrt{N}=1.3$ (b). We
  removed the parameter independent singularities at the endpoints, 
 $\widetilde{\delta \rho\ }(T)=\sqrt{T(1-T)}\delta \rho (T)$. The spin-orbit coupling  strength is characterized
  by $a_{\rm so}/\sqrt{N}=0.0\ ,0.3\ ,0.4\ ,10$ for solid, dashed, dotted and
  dot-dashed lines respectively.}
\label{fig:drho}
\end{figure}

In the crossover regime  the weak localization peaks in \eqref{eq:drhopure}
broaden, but the correction remains singular at
the endpoints. If all the $\gamma_m$-s are finite, the singularity is present
through a inverse square root factor proportional to
the $O(N)$ contribution. The second, nonsingular factor in \eqref{deltarho}
determines the form of the weak localization correction as the function of the
magnetic field and spin-orbit coupling.    In the absence of the
magnetic field ($\gamma_3=0$) this picture is modified by the inclusion of the
remanent Dirac delta contribution in place of the $\Gamma_3$ term. On Fig.  \ref{fig:drho} we illustrated the transition from weak localization
to weak antilocalization for two values of the magnetic field. The relevant
regions are close to the endpoints, where the peaks of the pure symmetry cases
\eqref{eq:drhopure} are located. Due to the antisymmetry of $\delta \rho (T)$ we
plotted only the part around $T=0$. 

The presence of spin-orbit coupling dependent and spin-orbit coupling independent
contributions is analogous to the case of higher dimensional systems studied
by Nazarov \cite{Nazarov95}, where they correspond to contributions coming from
spin-triplet and spin-singlet Cooperon
modes. The situation is very similar in our case. 
The basic building block of the diagrammatic expansion for $\delta \rho(T)$ is
the combination ${\mathcal T}C{\mathcal T}$, 
with $\mathcal{T}=\openone_2 \otimes \sigma_2\ $ and
\[C^{-1} = M \openone_2 \otimes \openone_2 - 
  \mbox{Tr}\, R \otimes \overline{R},  \]
where $\overline{()}$ denotes quaternion complex conjugation and the tensor product is defined with a backwards
multiplication:
\begin{equation}
  (\sigma_i \otimes \sigma_j) (\sigma_{i'} \otimes \sigma_{j'})
  = (\sigma_i \sigma_{i'}) \otimes (\sigma_{j'} \sigma_{j}).
\label{tensorrule}\end{equation}
The trace in the second term is understood as
\[\left (\mbox{Tr}\, R \otimes \overline{R}\right)_{\alpha \beta, \gamma \delta}=R_{ij, \alpha \beta}
\overline{R}_{ji,\gamma \delta},\]
where latin letters are channel indices, Greek letters refer to spin space and
summation over repeated indices is implied.
The very same structure emerges in the work of Brouwer et. al. \cite{BCH}, where the authors
identify $C$ as the equivalent of the Cooperon in the conventional
diagrammatic perturbation theory. 
In the limit $M\rightarrow \infty$ it becomes:
\begin{equation}C^{-1} = 2(N+x^2 +  a_{\rm so}^2) (\openone_2 \otimes \openone_2)    
      - a_{\rm so}^2 (\sigma_x \otimes \sigma_x +     \sigma_y \otimes \sigma_y).  \label{eq:C}\end{equation}
If according to the
multiplication rule \eqref{tensorrule} we define 
the action of a matrix on a vector as $(A v)_{\alpha
  \beta}=A_{\alpha \rho, \sigma \beta} v_{\rho \sigma}$, the
  spin-singlet and spin-triplet basis turns out to be the eigenbasis of the matrix ${\mathcal T}C{\mathcal T}$ with eigenvalues  \[\lambda^{-1}_{00}=2(N+\gamma_3),\
  \lambda^{-1}_{1\pm 1}=2(N+\gamma_1),\ \lambda^{-1}_{1   0}=2(N+\gamma_2).\]
As in Ref. \onlinecite{Nazarov95}, only the triplet eigenvalues
depend on the spin-orbit coupling strength.
The correction \eqref{deltarho} can be expressed as
\begin{equation}\begin{array}{c}\delta \rho(T) = \\
\quad\\
\frac{1}{4 \pi \sqrt{T(1-T)}}\sum\limits_{n=0,1}(-1)^n
  \left(\sum\limits_{m=-1}^1 \Gamma(\lambda_{1m},y_n)-\Gamma(\lambda_{00},y_n)
  \right),\end{array}\label{deltarholambda} \end{equation}
where the function $\Gamma$ is
\[ \Gamma(\lambda,y)=\frac{1-4N^2 \lambda^2}{(1-2N \lambda)^2+8N\lambda y}.\]
The appearance of the spin-orbit coupling
independent term is because of the decoupling of the singlet and triplet
sectors of  ${\mathcal T}C{\mathcal T}$.
Going back to the discussion of the weak localization - weak antilocalization transition with
\eqref{deltarholambda} in mind, we find that the weak localization peak is due to the
triplet terms with eigenvalue $\lambda_{1\pm 1}$, and the weak antilocalization peak comes from the singlet contribution.

An other property of the higher dimensional cases that persists also for quantum dots  is the breakdown of the
perturbation theory for small magnetic fields near $T=0,1$. More
precisely, for fields $x^2 \lesssim 1$  the applicability condition $\delta \rho
\ll \rho$ of the perturbation theory is violated in an interval of order
$O(x/N^2)$ from the endpoints of the spectrum. The conclusion is the same as
in Ref.~\onlinecite{Nazarov95}, namely, at small fluxes, for obtaining the
detailed  behaviour of the density near $T=0,1$ one has to treat the
problem in a nonperturbative way.

\section{Applications}
\label{sec:applications}
Having obtained the $\delta\rho(T)$, let us see some applications. In the
following we assume zero temperature. First we compute the 
conductance. We find 
\begin{equation}\begin{array}{ll}  G&=\frac{e^2}{h} \int_0^1 dT(\rho_0(T)+\delta \rho(T))T\\
&=\frac{e^2}{h} N\left[1-\left(\frac{1}{2(\gamma_1+N)}+\frac{1}{4(\gamma_2+N)}-\frac{1}{4(\gamma_3+N)}\right)\right] ,\end{array}\label{eq:deltaG}\end{equation}
where the second term represents the weak localization correction. It is
another verification of \eqref{deltarho}, as we recover the corresponding
result of Brouwer et. al. in Ref.~\onlinecite{BCH}. Note that the correction 
is of the form $\delta G \propto
\lambda_{00}-\sum_{m=-1}^1 \lambda_{1m}$ (Ref. \onlinecite{Hikami,AF}).  

As a second application we consider the shot noise power. We get
\[P=\frac{2e^3V}{h}\int_0^1 dT(\rho_0(T)+\delta
\rho(T))T(1-T)=\frac{2e^3V}{h}\frac{N}{4},\]
that is, the $O(1)$ contribution from $\delta \rho (T)$ is absent.
In the case of pure symmetry classes it is a known result, that the weak localization correction to the shot noise
power vanishes if the number of modes is the same in both leads
\cite{JPB,RMTQTR}. It was shown in Ref.~\onlinecite{Braun} that this persists to the case of
a $\beta=1 \rightarrow 2$ transition too. Our result allows us to extend this
prediction to the more general crossover interpolating between all of Dyson's
symmetry classes. The reason behind the absence of the $O(1)$ contribution is
that $T(1-T)$ is symmetric with respect to the point \mbox{$T=1/2$} and it is
integrated with the antisymmetric density function $\delta \rho(T)$.   

To see an example, where the weak localization correction is absent in the
limit of pure symmetry classes, but not in the crossover regime, let us
take the third cumulant of the distribution of the transmitted charge. It is the opposite of the shot noise in the sense,
that for cavities with  leads supporting the same number of channels the
$O(N)$ term vanishes \cite{blanter01,nagaev02-2},  thus 
the leading order of this quantity is determined by $\delta \rho$. The third cumulant is proportional to  
\[f_3=\int_0^1 dT(\rho_0(T)+\delta
\rho(T))T(1-T)(1-2T).\]
The weak localization correction trivially vanishes in the pure symmetry case
because of the factor $T(1-T)$ and the Dirac delta functions in
\eqref{eq:drhopure}. In the crossover regime we find
\[f_3=N\left(\frac{\gamma_1(\gamma_1+2N)}{16 (\gamma_1+N)^3}+\frac{\gamma_2(\gamma_2+2N)}{32 (\gamma_2+N)^3}-\frac{\gamma_3(\gamma_3+2N)}{32 (\gamma_3+N)^3}\right),\]
that is, the (ensemble average of the) third cumulant is ``crossover induced'' for a symmetric cavity. 
     
\section{Conclusions}
\label{sec:concl}
We investigated the crossover behaviour of the weak localization correction to
the density of the transmission eigenvalues between Dyson's three
symmetry classes $\beta=1,2,4$ for a case of a chaotic cavity with symmetric
leads. Using the stub model approach for the RMT description, with the help
of the diagrammatic method of Brouwer and Beenakker \cite{diagrams}, we carried out a
subleading order calculation in the small parameter $1/N$. Our main finding is 
a closed-form, analytical expression for the correction. 

We studied the weak localization - weak antilocalization transition in
detail. We found that the weak
\mbox{(anti)}localization peaks \eqref{eq:drhopure} of the case of pure symmetry classes broaden in the
crossover regime, but the correction still remains singular at the endpoint of
the spectrum. With our result \eqref{deltarho} at hand, we  gave a
quantitative description of the broadening and the crossover from localization
to antilocalization as the function of the magnetic field and spin-orbit coupling. 

We compared our results to the known cases of higher dimensionalities, and found
strong similarities. First, our result also splits into spin-singlet and spin-triplet parts, with only the triplet contribution depending on the spin-orbit
coupling. In the limits of pure symmetry classes, the weak localization peak 
comes from the triplet contribution, while the antilocalization peak is due to
the singlet part.
Second, we also find that for small magnetic fields, the perturbation theory fails to describe
the details of the density near the endpoints of the transmission
eigenvalue spectrum. 

We applied our
results to the conductance, the shot noise power and the third cumulant of the
distribution of the
transmitted charge. The conductance served as a test for our calculations, we
recovered the result of Ref.~\onlinecite{BCH} obtained in the framework of the
same model. For the shot noise power we found that the weak
localization correction is absent in the full crossover, due to the symmetry
of the transmission eigenvalue density.  For the third cumulant we found  opposite
behaviour. It is  crossover induced: the $O(N)$ term is absent, and the $O(1)$
contribution is
nonzero only in the crossover regime. 

Further directions of research could be to apply our result to obtain the weak
localization correction to the full statistics of the transmitted
charge. Another possibility would be
to extend our calculations to the case of cavities with asymmetric leads. In
that case, differently from the present results, we expect a nontrivial magnetic field and spin-orbit coupling
dependence also for the shot noise power. 

\section*{Acknowledgements}

We gratefully acknowledge discussions
with C. W. J. Beenakker.
This work is supported by E.~C.\ Contract No.~MRTN-CT-2003-504574.

\appendix
\section{Details of the calculation}
\label{app:details}
In this appendix we give the details of the derivation of our main result
\eqref{deltarho}. We adapt a procedure of  Brouwer and
Beenakker\cite{diagrams} that removes the nested geometric series in \eqref{eq:densgen},
appearing due to the inverse in the expression \eqref{eq:SU} for the S matrix. The price for this is the introduction
of more complicated matrix structures.

Let us introduce the $2M \times 2M$ matrices
\begin{equation}
\begin{array}{c}
\begin{array}{lll}
  {\bf S} = \left( \begin{array}{cc} S & 0\\ 0 & S^{\dagger} \end{array}
  \right), &  {\bf C} = \left( \begin{array}{cc} 0 & C_2 \\ C_1 & 0
  \end{array} \right), &  {\bf U} = \left( \begin{array}{cc} U & 0 \\ 0 & U^{\dagger}\end{array} \right),
\end{array}\\
\quad\\
\begin{array}{ll}
 {\bf F}(z) = \left( \begin{array}{cc} 0 & F'(z) \\
  F(z) & 0 \end{array} \right), & 
  {\bf R} = \left( \begin{array}{cc} R & 0\\ 0 & R^{\dagger} \end{array}
  \right),
\end{array}
\end{array}
\end{equation}
where for $S$ we use the representation \eqref{eq:SU} with $P$ being $M \times M$ matrix
\[ P_{ij} = 1\textrm{   if   } i=j \le 2N \textrm{   and   } 0 \textrm{
  otherwise,} \]
and the $M \times M$ matrices $C_1$ and $C_2$ are 
\begin{eqnarray*}
 (C_1)_{ij} &=&
1\textrm{   if   } i=j \le N \textrm{   and   } 0 \textrm{   otherwise}\\
C_2&=&P-C_1.\end{eqnarray*}
The Green functions $F_1(z)$ and $F_2(z)$ are defined as
\begin{subequations} \label{defGreen}
\begin{align}
  F(z) &=  C_1 (z - S C_2 S^{\dagger} C_1)^{-1}, \\
  F'(z) &= C_2 (z - S^{\dagger} C_1 S C_2)^{-1}.
\end{align}
\end{subequations}
The density of transmission eigenvalues can be obtained from $F(z)$ as 
\begin{equation}\rho(T) = -\pi^{-1}\lim_{\epsilon\rightarrow 0^+}\ \mbox{Im}\, \mbox{Tr}\,
\langle F(T + i \epsilon) \rangle.\label{Frho}\end{equation}

The matrix Green function ${\bf F}(z)$ can be expressed as
\begin{equation}
\begin{array}{c}
  {\bf F}(z) = \\
\quad\\
    (2 z)^{-1} \sum_{\pm}
    \left( {\bf C} \pm {\bf C}
    \left[1 - {\bf U}({\bf R} \pm {\bf C} z^{-1/2})
    \right]^{-1} {\bf U}  {\bf C} z^{-1/2} \right)   \\
\quad\\  
=  (2z)^{-1} \sum_{\pm}
    \left[ {\bf C} \pm {\bf A}_{\pm}
     ({\bf F}_{\pm} - {\bf X}_{\pm}) {\bf B}_{\pm} 
    \right],
\end{array} \label{eq:Fcalres1}
\end{equation}
with ${\bf X}_{\pm} = {\bf R} \pm {\bf C} z^{-1/2}$ and ${\bf F}_{\pm} = {\bf
  X}_{\pm} (1 - {\bf U} {\bf X}_{\pm})^{-1} $. We defined ${\bf A}_{\pm}$ and ${\bf B}_{\pm}$ such that $  {\bf A}_{\pm} {\bf X}_{\pm} = {\bf C} $, $  {\bf X}_{\pm} {\bf B}_{\pm} = {\bf C} z^{-1/2}$.

To get the ensemble average of ${\bf F}$, one has to calculate the COE average of ${\bf
  F}_{\pm}$. In the following ${\bf  F}_{\pm}$ refers
  to this unitary average.  It  is related to
  the self energy ${\bf \Sigma}_{\pm}$ through the Dyson equation
\begin{equation} \label{eq:Dyson}
  {\bf F}_{\pm} =
    {\bf X}_{\pm}\left( 1 + {\bf \Sigma}_{\pm} {\bf F}_{\pm} \right), \ \
\end{equation}  
We can express $\langle {\bf F} \rangle$ directly
through  ${\bf \Sigma}_{\pm}$ as
\begin{equation} \label{eq:calfeq}
  \langle {\bf F} \rangle = (2 z)^{-1} \sum_{\pm} \left( {\bf C} \pm {\bf C}
  \left \langle(1 - {\bf \Sigma}_{\pm} {\bf X}_{\pm})^{-1} {\bf
  \Sigma}_{\pm}\right \rangle {\bf C} z^{-1/2} \right)\end{equation}

First we calculate ${\bf F}_{\pm}$ to leading order
  in $\frac{1}{N}$. To this order we have to consider the planar diagrams
  only. Denoting the resulting series as ${\bf F}^{(0)}_{\pm}$,
for the self energy we find  
\begin{equation}
{\bf \Sigma}^{(0)}_{\pm} =
    \sum_{n=1}^{\infty} W_n \left( {\cal P} {\bf F}^{(0)}_{\pm} \right)^{2n - 1},
\label{planarsigma}\end{equation}
where the coefficients $W_n$ are given as \cite{diagrams}
\[W_n = {1 \over n} N^{1-2 n} (-1)^{n-1} {2 n - 2 \choose n-1}.\]
The operator ${\cal P}$ acts on a $2M \times 2M$ matrix ${\bf A}$ as
\begin{equation*}
  {\bf A} = \left( \begin{array}{cc}
    A_{11} & A_{12} \\ A_{21} & A_{22} \end{array}\right), \ \
  {\cal P} {\bf A} = \left( \begin{array}{cc}
    0 &  \mbox{tr}\, A_{12} \\  \mbox{tr}\, A_{21} & 0
  \end{array} \right).
\end{equation*}
With the help of the generating function
\[h(z) = \sum_{n=1}^{\infty} W_{n} z^{n-1} = {1 \over 2 z} \left( \sqrt{M^2 +
  4 z} - M \right)\]
we can write equation \eqref{planarsigma} as
\begin{equation*}
\begin{array}{c}
  {\bf \Sigma}^{(0)}_{\pm} =\\
\quad\\ 
\left( {\cal P} {\bf X}_{\pm} (1 - {\bf \Sigma}^{(0)}_{\pm} {\bf X}_{\pm})^{-1}\right)\, h \left( \left({\cal P} {\bf X}_{\pm} (1 - {\bf \Sigma}^{(0)}_{\pm} {\bf X}_{\pm})^{-1} \right)^{2} \right).
\end{array}
\end{equation*}

The solution  is
\begin{equation}
  {\bf \Sigma}_{\pm}^{(0)} = \pm \left(\sqrt{z} - \sqrt{z-1}\right) \left( \begin{array}{cc} 0 & 1 \\ 1 & 0 \end{array} \right).
\label{eq:sigma0}\end{equation}

From \eqref{eq:sigma0} it follows that
\[\mathrm{Tr}F_0(z)=\mathrm{Tr}F_0'(z)=\frac{2N}{\sqrt{z(z-1)}},\]
from which we get the well known result \eqref{rho0} for the density of transmission
eigenvalues.

In accounting for the weak localization correction let us write the self
energy as
\begin{equation}{\bf \Sigma}_{\pm}={\bf \Sigma}_{\pm}^{(0)}+\delta{\bf \Sigma}_{\pm}.\end{equation}
It follows from \eqref{eq:calfeq}, that ${\bf F}$ splits up too, 
\[{\bf F}={\bf F}^{(0)}+\delta{\bf F}, \]
with $\delta{\bf F}$ containing the weak localization correction to the Green functions
\eqref{defGreen} in its off-diagonal blocks.
Up to first order in $\delta{\bf  \Sigma}_{\pm}$, after a little algebra we get  
\begin{equation}\delta{\bf
    F}=\frac{1}{2}\left(\frac{1}{z}-\frac{1}{z-1}\right)\sum_{\pm} {\mp}\sqrt{z}\ {\bf
    C}\ \delta{\bf \Sigma}_{\pm} {\bf C}.\label{dFdsigma}\end{equation}

The contributions to the self  energy correction $\delta{\bf \Sigma}_{\pm}$
come from the $O(\frac{N^\alpha}{M^{\alpha+1}})$ terms in the large-$M$
expansion of ${\bf \Sigma}_{\pm}$. These can be sorted as
\begin{equation}\delta{\bf \Sigma}_{\pm}=\delta{\bf
    \Sigma}^{(e)}_{\pm}+\sum_{n=1}^{\infty} W_n \delta\left( {\cal P} {\bf
    F}_{\pm} \right)^{2n - 1}.\label{deltasigmasc}\end{equation}

\begin{figure}
\epsfig{file=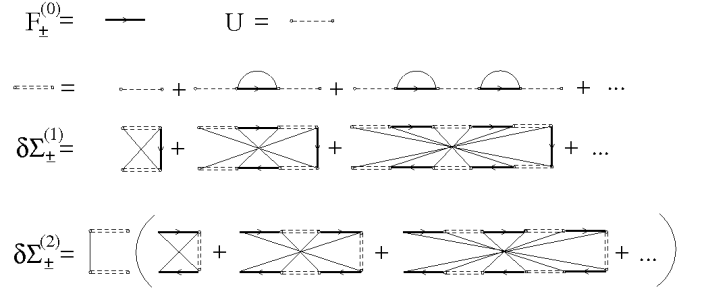,height=4cm}
\caption{Diagrams contributing to the explicit part 
    $\delta {\bf \Sigma}^{(e)}_{\pm}$ of the weak localization correction to the self energy.}
\label{fig:diag}
\end{figure}

The first term consists of diagrams with the outermost $U$-cycle being
    non-planar (see \mbox{Fig. \ref{fig:diag}}), and a term due to the sub-leading order in the
    large-$M$ expansion of the cumulant coefficients
\begin{equation}\delta{\bf  \Sigma}^{(e)}_{\pm}=\delta{\bf
    \Sigma}^{(1)}_{\pm}+\delta{\bf  \Sigma}^{(2)}_{\pm}+\underbrace{\sum_{n=1}^{\infty}
    \delta W_n \left( {\cal P} {\bf F}^{(0)}_{\pm} \right)^{2n -
    1}}_{\delta{\bf  \Sigma}^{(\delta W)}_{\pm}},\label{sigmae}\end{equation}
with
\(\delta W_n=-\frac{(-4)^{n-1}}{M^{2n}}\).
Evaluating the diagrams of \mbox{Fig. \ref{fig:diag}} for $\delta{\bf
    \Sigma}^{(1)}_{\pm}$ we find  
\begin{equation}
\begin{array}{c}
\delta{\bf  \Sigma}^{(1)}_{\pm}=\\
\quad\\
\left(\begin{array}{cc}
aE_{\alpha \sigma,\alpha' \sigma'}R^*_{\sigma' \sigma} & bG_{\alpha
  \sigma,\alpha' \sigma'}\left(R^\dagger R+2C_2\right)_{\sigma' \sigma}  \\
bG_{\alpha \sigma,\alpha' \sigma'}\left(R^\dagger R+2C_1\right)_{\sigma' \sigma} & aE_{\alpha \sigma,\alpha' \sigma'}R^T_{\sigma' \sigma} \\
\end{array} \right)
\end{array}
\end{equation}
where $()^*$ denotes the complex conjugate, Greek indices refer to spin space and we assumed summation for repeated
indices. Furthermore
\[a=\frac{\sqrt{z}+\sqrt{z-1}}{2\sqrt{z-1}},\quad b=\pm \frac{1}{2\sqrt{z-1}}, \]
\begin{equation}
E=-2Nb^2\ \mathcal{T} \Pi  \mathcal{T}, \quad
G=a^2\ \mathcal{T}\ C^{-1}  \Pi \ \mathcal{T},
\end{equation}
with 
\[\Pi=\left(a^4 C^{-2}-(2Nb^2)^2\right)^{-1}.\]
The matrices $\mathcal{T}$ and $C$ are defined as in Sec. \ref{sec:result}.

The second term $\delta{\bf  \Sigma}^{(2)}_{\pm}$ is
\begin{equation}
\delta{\bf  \Sigma}^{(2)}_{\pm}=
\left((s_1+s_2) Q_{11}+2s_1 Q_{12}\right)_{\alpha \sigma,\alpha'\sigma}  \left(\begin{array}{cc}
0 & 1 \\
1 & 0
\end{array} \right) ,
\end{equation}
where
\[
\begin{array}{l}Q=M\left(\begin{array}{cc}a^2 G+b^2 E\  & a^2 E+b^2 G\\ a^2 E+b^2 G\  & a^2
  G+b^2 E\end{array} \right)\times \\
\quad\\
\left(\begin{array}{cc} a^2 {\rm Tr}R \otimes R^* \ & b^2
  (M-2N)\\ b^2 (M-2N)\  & a^2 {\rm Tr}R \otimes R^*,\end{array} \right)
\end{array}\]
where the trace is defined as in Sec.~\ref{sec:result} and 
\[ s_1=-\frac{b}{M^2}\left(\frac{z-1}{z}\right)^{3/2}\quad s_2=\left(1-4z\frac{b^2}{a^2}\right)s_1.\]
Doing the summation in the third term in \eqref{sigmae} we get  
\begin{equation}\delta{\bf  \Sigma}^{(\delta W)}_{\pm}=-\frac{b}{M}\frac{z-1}{z}\left(\begin{array}{cc}
0 & 1\\
1 & 0
\end{array} \right) .\end{equation}

The second term in \eqref{deltasigmasc} contains sub-leading order diagrams,
that have planar outermost $U$-cycles. Up to first order in $\delta{\bf
  \Sigma}_{\pm}$ 
\[\begin{array}{c}\delta\left( {\cal P} {\bf F}_{\pm} \right)^{2n - 1}=\\
\quad\\
 \left( {\cal P} {\bf  F}^{(0)}_{\pm}+{\cal P}\left( {\bf F}^{(0)}_{\pm} \delta{\bf
  \Sigma}_{\pm}
{\bf F}^{(0)}_{\pm}\right) \right)^{2n - 1}
 -\left( {\cal P} {\bf F}^{(0)}_{\pm} \right)^{2n - 1}.\end{array}\]
 Putting everything together we see, that \eqref{deltasigmasc} is a
  (linear) self-consistency equation for $\delta{\bf  \Sigma}_{\pm}$, which can
  be solved straightforwardly, if from \eqref{dFdsigma} we notice, that for the transmission
  eigenvalue density it is enough to get ${\rm Tr}_2 \ \delta{\bf
  \Sigma}_{\pm}$, where we denoted the spin-trace as ${\rm Tr}_2$. 
Substituting the solution in \eqref{dFdsigma} in the lower left block we get the weak localization
 correction to $F(z)$, from which using  \eqref{Frho} we arrive to the result \eqref{deltarho}.


\begin{thebibliography}{99}
\bibitem{tauso}Regarding the relation between the spin
orbit time $\tau_{\rm so}$ and the Rashba/Dresselhaus parameters we refer to
Refs. \onlinecite{BCH,CBF}.
\end{thebibliography}
\end{document}